\documentclass[a4paper]{jpconf}
\usepackage{graphicx}
\usepackage[utf8]{inputenc}
\usepackage{braket}
\usepackage{verbatim}
\usepackage{subcaption}
\usepackage{hyperref}
\usepackage{float}
\usepackage{amsmath}
\begin{document}
\title{Landauer current and mutual information in a bosonic quantum dot}

\author{Hrushikesh Shashikant Sable}
\address{Physical Research Laboratory, Ahmedabad, 380009, India}
\ead{hsable@prl.res.in}
\author{Devendra Singh Bhakuni}
\address{Indian Institute of Science Education and Research Bhopal, M.P, 462066, India}
\ead{devendra123@iiserb.ac.in}

\author{Auditya Sharma}
\address{Indian Institute of Science Education and Research Bhopal, M.P, 462066, India}
\ead{auditya@iiserb.ac.in}

\begin{abstract}
We study the quantum transport of bosons through a quantum dot coupled to two macroscopic heat baths $L$ and $R$, held at fixed temperatures $T_{L}$ and $T_{R}$ respectively. We manage to cast the particle as well as the heat current into the Landauer form. Following the correlation matrix approach, we compute the time-dependent mutual information of the dot with the baths. We find that mutual information goes logarithmically as the number of bosons, and at low temperatures, it is possible to set up the parameters in such a way that in steady-state, the mutual information goes quadratically as a function of current.
%We study the mutual information of a quantum dot coupled to two macroscopic heat baths $L$ and $R$, consists of bosons and held at fixed temperatures $T_{L}$ and $T_{R}$ respectively. We compute the particle as well as the heat current flowing into the dot and managed to recast it into Landauer form. The relationship between the steady state value of mutual information and the current is also established .
%This work is a generalization of the work done by (Sharma Auditya, and Eran Rabani. Physical Review B 91.8 (2015): 085121) but for the bosonic baths.
\end{abstract}

\section{Introduction}
Quantum entanglement~\cite{mintert2009basic} has proven to be a useful quantity to probe in a variety of phenomena. The last decade or so has seen a great proliferation of activity 
at the interface between quantum information and condensed matter physics~\cite{amico2008entanglement,laflorencie2016quantum}, two disciplines that have been traditionally considered to be distinct and rather far apart. One specific context
in which this merger has shown itself to be particularly interesting, is that of quantum transport. Although the Landauer approach to quantum transport~\cite{Haug2008Quantum}~\cite{RLandauer1970Quantumtransport}~\cite{Buttikercurrent} has been around for many decades, and has been developed extensively, only recently~\cite{auditya2015landauer} has the connection between current through a quantum dot, and the quantum correlations that develop between the dot and the leads in a nonequilibrium setting, been shown to be intimate. 
A motivation for the current study is to explore how this connection plays out when the statistics involved is bosonic, rather fermionic. When bosonic degrees of freedom are involved, in addition to the particle current, a \emph{heat current} also becomes relevant~\cite{dhar2006heat,DSegalANitzanHeattrasport2003,heatcurrent}. The drive for the dynamics is generated by a temperature gradient between the baths.

When quadratic Hamiltonians are involved, Wick's theorem can be exploited to provide a prescription for computing entanglement in the eigenstates in terms of the eigenvalues of an underlying correlation matrix~\cite{peschel2012special,peschel2003RDM}; this approach has been successfully employed in a variety of contexts. However, the literature on entanglement in nonequilibrium phenomena~\cite{eisler2014area}, and particularly for bosonic sytems~\cite{eisler2014entanglement}, is relatively sparse. A nonequilibrium system typically involves mixed states, where entanglement is rather hard to study. Mutual information,~\cite{mutualinfo1,mutualinfo2,mutualinfo3} which includes both classical and quantum correlations is one of the commonly studied quantities in such a scenario, and we too adopt this for our study. 

The layout of the current paper is as follows. In the next section, we
describe our model, and the Ohmic bath spectral density used, from
which the parameters are obtained.  In the
following section, we write down the exact non-equilibrium density matrices
for the dot and the baths.  Introducing a new set of bosonic
operators, we calculate the current in the subsequent section.  We
learn that expressions for the particle current calculated and heat
current calculated are very similar, and can be recast in the Landauer
form. Next, exploiting the fact that the reduced density matrices have
a thermal form at all times in the dynamics, the spectra of the
reduced density matrices can be related to the time-dependent
correlation matrices. We work out the expression for the von Neumann
entropy of the subsystems and calculate the mutual information. A results and discussion section then 
collects and compares current and mutual information. The last section is reserved for conclusions and summary.

\section{Model Hamiltonian} \label{Model hamiltonian}
The model consists of a quantum dot in the center coupled to two bosonic baths - left $L$ and right $R$. The Hamiltonian can be written as 
\begin{equation} \label{eq1}
\begin{split}
H &= H_{L} + H_{R} + H_{D} + H_{LD} + H_{RD} \\ 
H_{D} &= \epsilon_{d} d^{\dagger}d \\
H_{L,(R)} &= \sum_{k\in L(R)}^{}\epsilon_{k} c_{k}^{\dagger} c_{k} \\
H_{LD,(RD)} &= \sum_{k\in L(R)}^{}t_{k}(c_{k}^{\dagger}d + c_{k}d^{\dagger}),
\end{split}
\end{equation}
where $\epsilon_{d}$ is energy of the dot, $\epsilon_{k}$ is the $k^{\text{th}}$ mode energy of a bath and $t_{k}$ is the coupling of the dot and the $k^{\text{th}}$ mode. Throughout the paper we set $\hbar = 1, e= 1$ and $k_B=1$.
One can obtain the couplings $t_k$ from the spectral density function $J(\omega)$, for which a general expression is~\cite{breuer2002theory}
\begin{equation} 
J(\omega)=\sum_{k}^{N} t^{2}_{k} \delta(\omega - \omega_{k}).
\end{equation}
Following the standard approach for bosons~\cite{ohmicdensity}, we model the baths using the Ohmic spectral density with an exponential cut-off $\omega_{c}$ at both ends $(L,R)$~\cite{weiss2012quantum}:
\begin{equation} \label{eq3}
J(\omega) = \eta\omega e^{\frac{-\omega}{\omega_{c}}},\qquad\quad
\end{equation}
where $\eta$ is the damping/friction constant.  Integrating in a small $\Delta\omega$ window around $\omega_{k}$, we obtain
\begin{equation}
t_{k}=\sqrt{\eta\omega_{k}e^{\frac{-\omega_{k}}{\omega_{c}}}\Delta\omega}.
\end{equation}
 \section{Non-equilibrium density matrix} \label{Non-Equilibrium Density matrix}
Initially the left and right baths are separately in equilibrium, each at zero chemical potential and different temperatures $T_{L}$ and $T_{R}$ respectively.
So the initial state of the system can be described by
\begin{equation}
\rho(0) = \rho_{L}(0)\otimes\rho_{R}(0)\otimes\rho_{D}(0),
\end{equation}
where the left and the right baths are in thermal equilibrium:
\begin{equation}
\rho_{L}(0) = \frac{\exp(-\beta_{L}\hat{H_{L}})}{Z_{L}},\quad \rho_{R}(0) = \frac{\exp(-\beta_{R}\hat{H_{R}})}{Z_{R}},
\end{equation}
and the dot's state is represented as $\rho_{D}(0) = n_{0}d^{\dagger}d + (1 - n_{0})dd^{\dagger}$.
At $t=0$, the coupling Hamiltonian $H_{LD,(RD)}$ is turned on and the unitary time evolution of the density matrix $\rho(t) = e^{- i H t}\rho(0)e^{i H t}$  is governed by the Hamiltonian in Eq.~\ref{eq1}. 
%\begin{equation}
%H = \sum_{k=1}^{N} \epsilon_{k} c_{k}^{\dagger}c_{k} + \sum_{i=1}^{N-1} t_{i}(c_{i}^{\dagger}c_{N} + c_{N}^{\dagger}c_{i}),
%\end{equation}
The Hamiltonian becomes diagonal when expressed in terms of new
bosonic operators $a_{\alpha} = \sum_{i=1}^{N}\psi_{\alpha}(i)c_{i}$,
where $\psi(i)$ is the $i^{th}$ eigenvector of $H$ corresponding to
eigenvalue $e_i$~\cite{auditya2015landauer}, where $N$ is the total
number of levels including the baths and the dot, $N = N_{L}+N_{R}+1$,
$N_{L}$ and $N_{R}$ being the energy-levels in the left and right
baths respectively. The $N^{\text{th}}$ index is used for the dot.

\section{Current in Landauer form} 
The left(right) particle current is the rate of change of occupancy of left (right) bath.
\begin{equation}
I_{L/R}(t) = - \frac{d \langle \hat{N}_{L/R} \rangle}{dt} = i\sum_{i \in L(R)}t_{i} \big\langle(c_{i}(t)c_{N}^{\dagger}(t) - c_{i}^{\dagger}(t)c_{N}(t)) \big \rangle. 
\end{equation}
The overall current is defined as $I(t) = \frac{I_{L}(t) - I_{R}(t)}{2}$ which simplifies to~\cite{auditya2015landauer}
\begin{equation} \label{eq.current}
I(t) = \sum_{k} t^{'}_{k}\sum_{i,j,p=1}^{N}\mbox{Im}\Big(\psi_{i}(N)\psi_{j}^{*}(k)\psi_{i}^{*}(p)\psi_{j}(p)e^{i(e_{i}-e_{j})t}\Big) f_p,
\end{equation}
where $t^{'}_{k} = t_{k}$ if $k \in L$ and $t^{'}_{k} = -t_{k}$ if $k \in R$ and $f_p$ is the Bose Einstein function $f(\epsilon_{p},T_{L})$ for modes on the left bath, $f(\epsilon_{p},T_{R})$ for modes on the right bath and $n_{0}$ for the dot population.\\
The Landauer form of the current is~\cite{MeirCurrent} 
\begin{equation}
I(E) = \int_{}^{} \mathcal{T}(E) \Big(f_{L}(E) - f_{R}(E)\Big) dE,
\end{equation}
where the $\mathcal{T}(E)$ is the transmission function and $f_{L}$ and $f_{R}$ are the distribution functions for the left and the right bath respectively.  The symmetry present in the energy levels and the coupling constants of the baths allows us to express the current in the Landauer form as 
\begin{align}\label{Landauerform}
I(t) = 2\sum_{p=1}^{N_{L}}\sum_{k=1}^{N_{L}} t_{k}\sum_{i=1,3,.;j=2,4.}^{N}\mbox{Im}\Big( \psi_{i}(N)\psi_{j}^{*}(k)\psi_{i}^{*}(p)\psi_{j}(p)e^{i(e_{i}-e_{j})t}\Big)\Big(f_{B}(T_{L}) - f_{B}(T_{R})\Big).
\end{align}
In analogy with the particle current, one can work out the expression for the heat current as
\begin{equation}
J_{L/R}(t) = - \frac{d \langle \hat{H}_{L/R} \rangle}{dt} = i\sum_{i \in L(R)}t_{i}\epsilon_{i}\big \langle(c_{i}(t)c_{N}^{\dagger}(t) - c_{i}^{\dagger}(t)c_{N}(t)) \big \rangle. 
\end{equation}
The expression for heat current has the same form as the particle current:
\begin{equation} \label{eq.heat current}
J(t) = \sum_{k} t^{'}_{k}\epsilon_{k}\sum_{i,j,p=1}^{N}\text{Im} \Big( \psi_{i}(N)\psi_{j}^{*}(k)\psi_{i}^{*}(p)\psi_{j}(p)e^{i (e_{i}-e_{j})t} \Big)f_{p}.
\end{equation}
One can also recast the heat current into Landauer form as in Eq.\ref{Landauerform} with $t_k$ replaced by $t_k \epsilon_k$.
\section{Mutual information} \label{Mutual information formalism}
We wish to study the evolution of the quantum correlations between the quantum dot and the baths. Since the initial state is mixed, von Neumann entropy is unsuitable as a measure of entanglement. Therefore we study mutual information which measures the total correlations: quantum and classical.
The mutual information between the dot and the baths is defined as
\begin{equation} \label{eq: 3.15}
S = S_{D} + S_{LR} - S_{\mbox{full}} ,
\end{equation} 
where $S_{D} = -\text{Tr}(\rho_{D}\mbox{ln}\rho_{D})$, $S_{LR} = -\text{Tr}(\rho_{LR}\mbox{ln}\rho_{LR})$ and $S_{\text{full}} = -\text{Tr}(\rho\mbox{ln}\rho)$ corresponds to the von Neumann entropies of the dot, baths and the full system respectively.
It can be shown that the reduced density matrices have a thermal-like form at all times in the dynamics~\cite{auditya2015landauer}. This fact allows for the characterization of the reduced density matrix of a subsystem by the single particle correlators only~\cite{peschel2003RDM}. Then the von Neumann entropy $S_{G}$ of a subspace $G$ is given as~\cite{peschel2012special,ding2009entanglement}%Frerot2015,
\begin{equation} \label{eq: 3.16}
S_{G} = \sum_{\sigma = 1}^{N_{G}}[(1 + C_{\sigma})\mbox{ln}(1 + C_{\sigma}) - C_{\sigma}\mbox{ln}(C_{\sigma})],
\end{equation}
where $C_{\sigma}$ are the eigenvalues of the correlation matrix defined within subspace $G$ and $N_{G}$ is the total number of sites in the subspace. The above formula is also valid for a time dependent correlation matrix. Thus to calculate mutual information, we need to calculate the correlation matrix of the baths and the dot at each time step and diagonalize it to obtain the eigenvalues. 

\section{Results and discussion}
We now present numerical results for a large but finite system with $N_{L}$=128, $N_{R}$=128 sites in the left and the right bath respectively. We have verified that this system size is large enough to have converged to the thermodynamic limit (by the indistinguishability of data obtained for all the quantities from a system of half the size). To generate the spectral density defined in Eqn.~\ref{eq3}, we choose the cut-off $\omega_c=20$ and discretization $\Delta \equiv \omega_c/N_L$. The energy levels in each of the baths range uniformly starting from a small positive value of $\Delta$ upto $\omega_c$ in steps of $\Delta$.
\begin{figure}[!h]
\begin{minipage}{11.8pc}
\includegraphics[width=9pc,angle=-90]{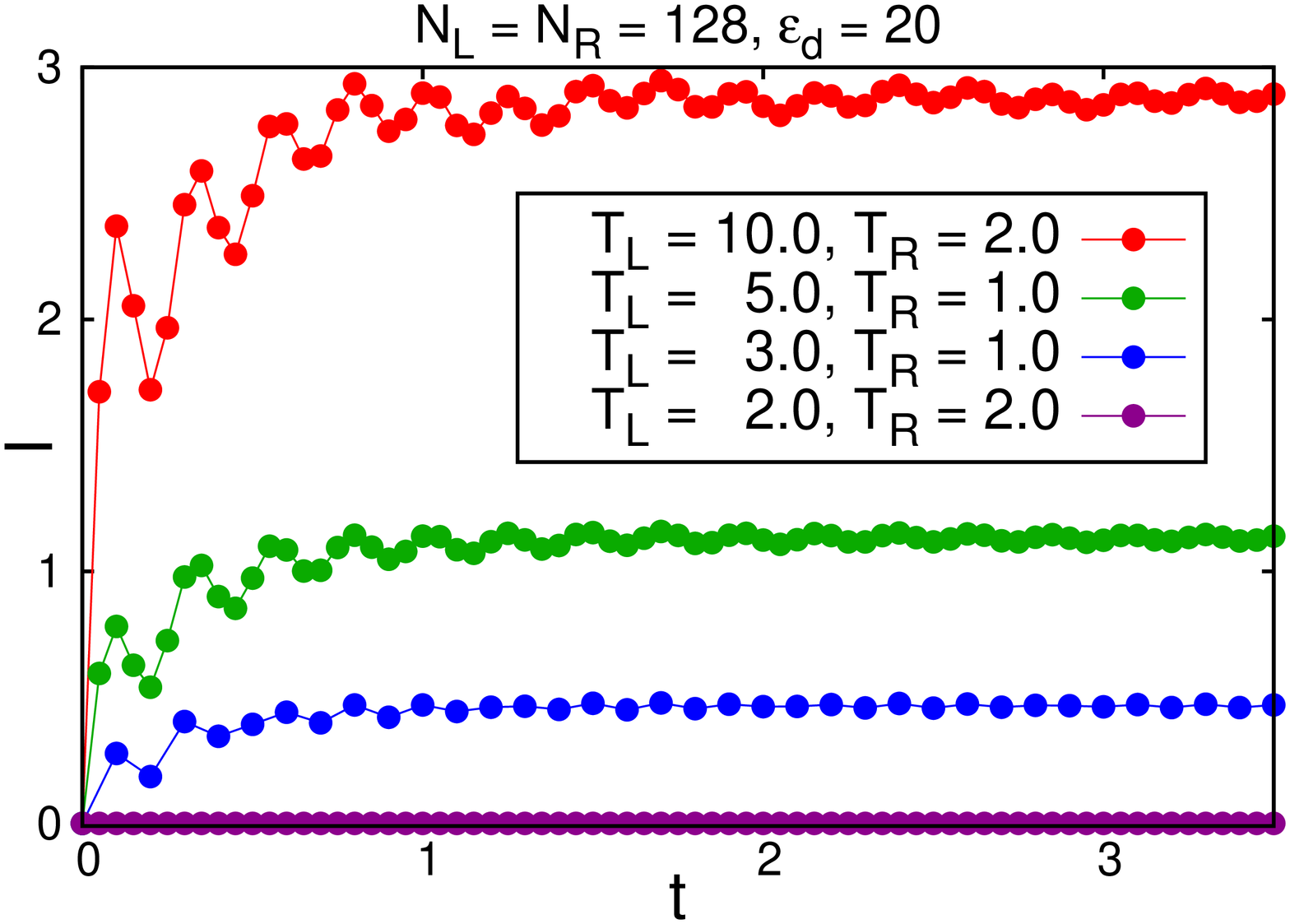}
%\caption*{\label{label1}Current vs Time.}
%\caption*{A}
\end{minipage}\hspace{0.5pc}
\begin{minipage}{11.8pc}
\includegraphics[width=9pc,angle=-90]{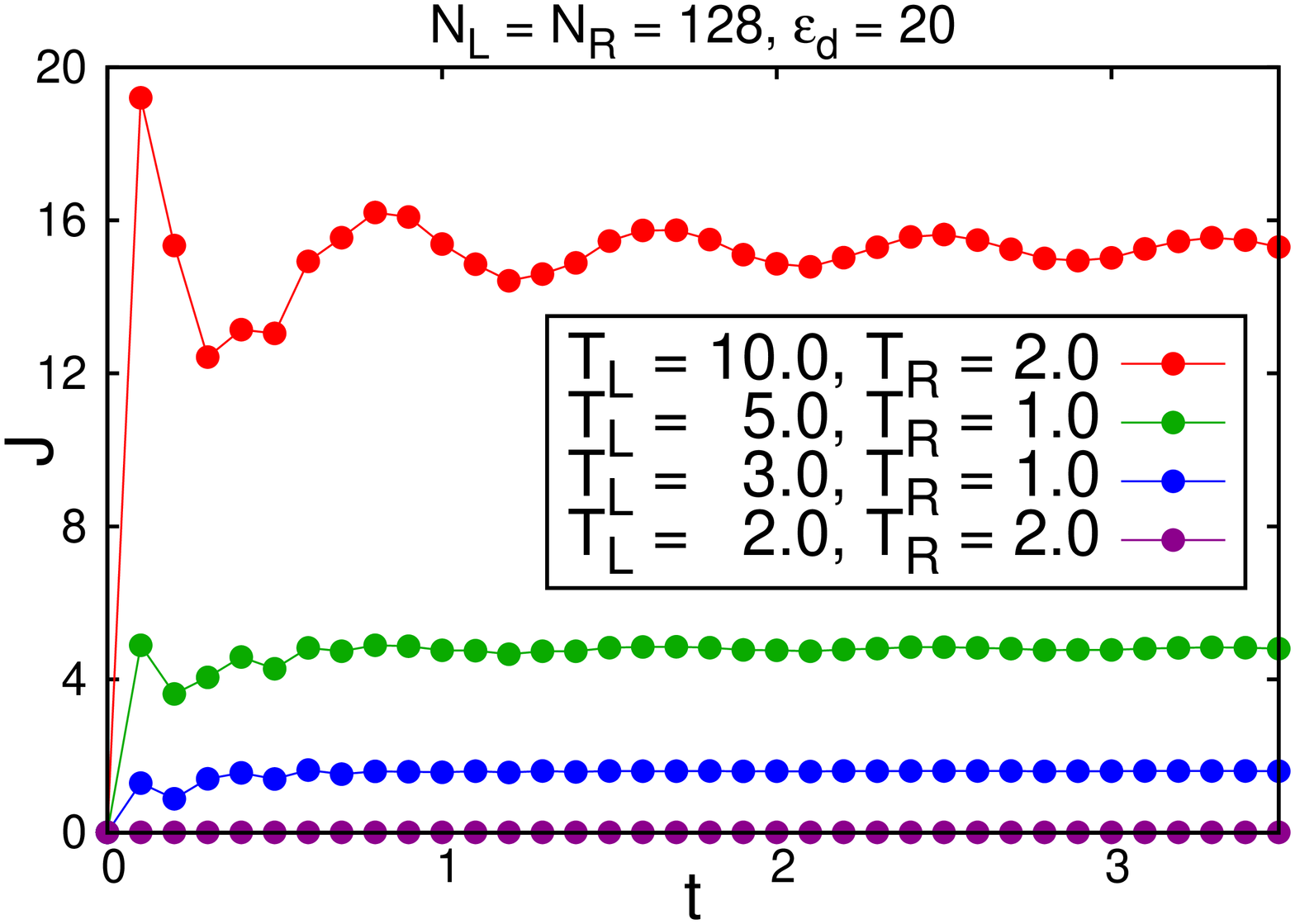}
%\caption*{\label{label2}Heat current vs Time.}
%\caption*{B}
%\vspace{-0.5cm}
\end{minipage} \hspace{0.2pc}
\begin{minipage}{11.8pc}
\includegraphics[width=9pc,angle=-90]{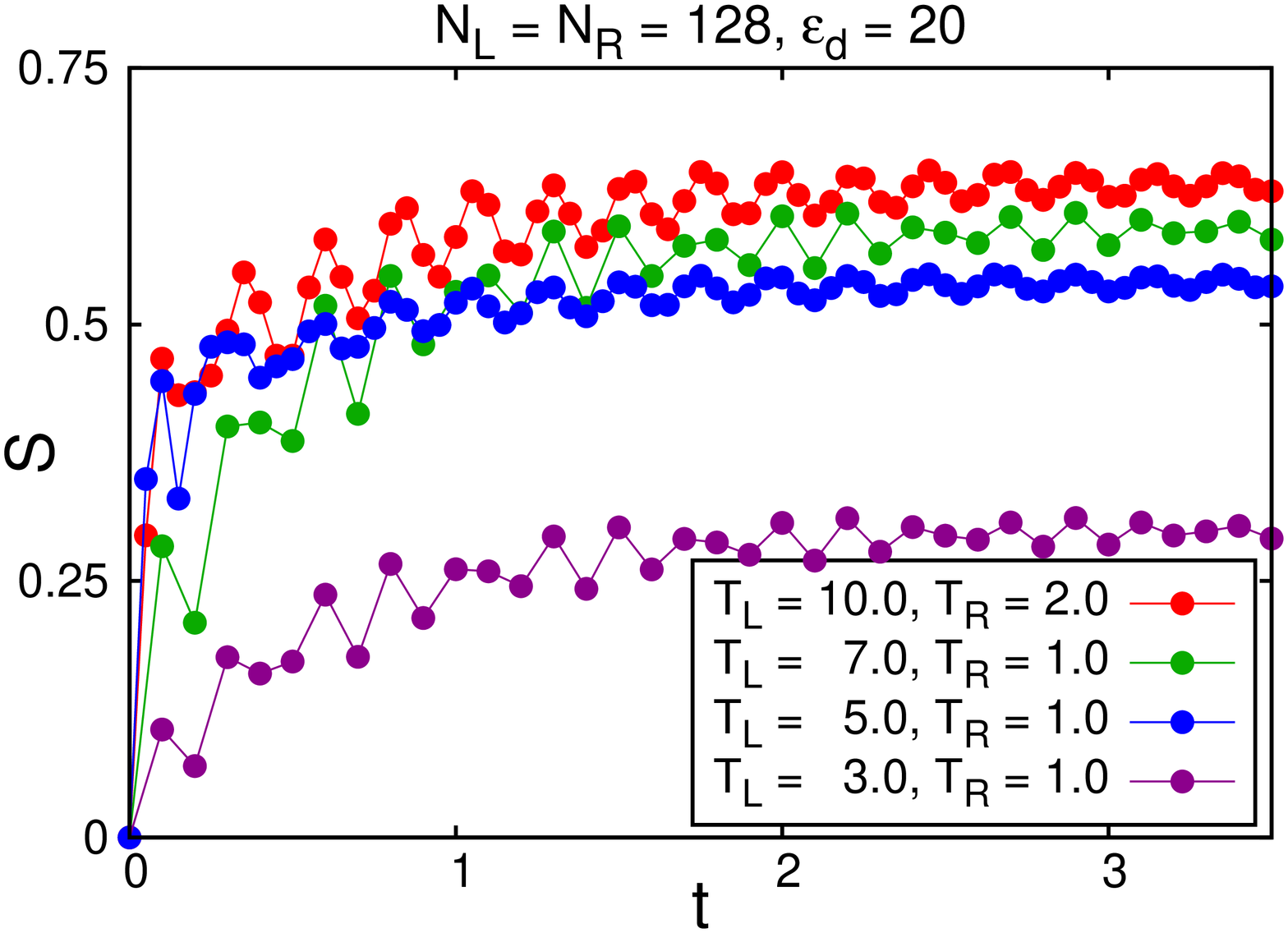}
%\caption*{\label{label3}Mutual Information vs Time.}
%\caption*{C}
%\vspace{-0.5cm}
\end{minipage} 
\caption{\label{label3} Variation of particle current I, heat current J and mutual information S with time. A nonequilibrium steady state is attained at long times.}
\end{figure}

The following quantities: particle current $I$, heat current $J$ and mutual information $S$ are plotted in Fig.~\ref{label3} as a function of time. We see a close relationship between $I$ and $S$ as well as $J$ and $S$. Greater the current is, greater are the correlations between the baths and the dot. Thus non-equilibrium current works as a physical observable to measure the correlations in the system~\cite{auditya2015landauer}. % Thus in an experiment, one can tune in the parameters to increase the correlations to obtain high current. 
 Since the transient behaviour of the current and the mutual information involves rapid changes, we focus on their steady state values. We define a dimensionless ratio $r = \frac{T_{L}-T_{R}}{T_{L}+T_{R}}$ and plot the steady state value of the current $I_\infty$ and the mutual information $S_\infty$ for different ratios $r$.

\begin{figure}[t]
\begin{minipage}{14.2pc}
\includegraphics[width=17.2pc,height=10.5pc]{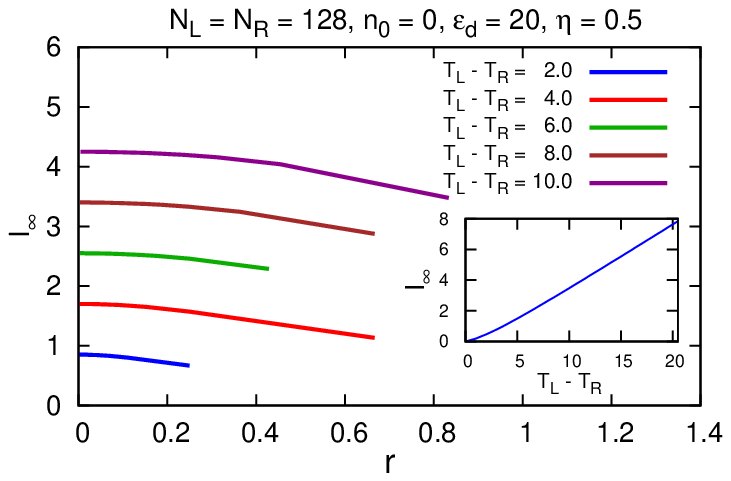}
\end{minipage}\hspace{3pc}
\begin{minipage}{14.5pc}
\includegraphics[width=17.2pc,height=10.5pc]{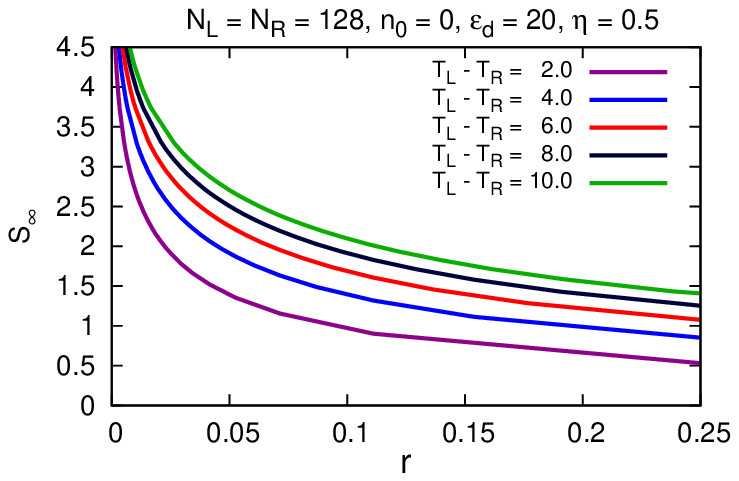}
\end{minipage} 
\caption{\label{label5} The steady state value of current and mutual information plotted against the  ratio $r$ for various temperature differences.
% The inset shows $I_{\infty}$ as a function of $T_{L}-T_{R}$.} 
The steady state value of current only depends on the temperature differences as depicted by the inset. $S_\infty$ decreases on increasing $r$, or increases on increasing both $T_L$ and $T_R$. }
\end{figure}
From Fig.~\ref{label5}, we see that the current depends only on the
temperature difference $T_{L}-T_{R}$ at high temperatures,
which is a direct consequence of the Landauer formula. We also observe that $S_{\infty}$ increases on increasing both $T_{L}$ and $T_{R}$.
\begin{figure}[h]
\includegraphics[width=15pc]{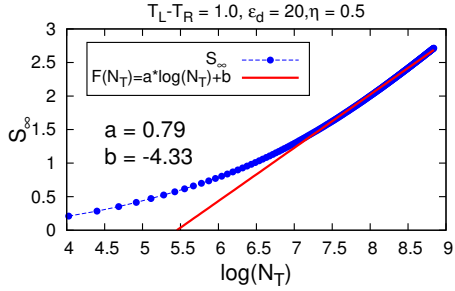}\hspace{1pc}
\begin{minipage}[b]{20pc}\caption{\label{labell}%Figure shows the variation of mutual information with the number of bosons\\ 
A logarithmic behaviour of mutual information with the total number of bosons $N_{T}$ is obtained for a large $N_{T}$. The total number of bosons $N_{T}$ can be calculated from the distribution function as  
$N_{T} = \sum_{i=1}^{N_{L}}f_{B}(\epsilon_{i},\mu,T_{L})+\sum_{i=1}^{N_{R}}f_{B}(\epsilon_{i},\mu,T_{R})+n_{0}$.}
%A logarithmic behaviour is obtained for large $N_T$.}
% An excellent fit to the form $a\log(N_{T}+b)+c$ can be obtained for large $N_T$. }
\end{minipage}
\end{figure} 
Higher temperatures $T_{L}$ and $T_{R}$ correspond to a larger total number of bosons $N_{T}$ in the left and the right baths respectively and hence greater the correlations between the baths and the dot.
Fig.~\ref{labell} shows that $S_{\infty}$ is found to be logarithmic with respect to $N_T$ for large $N_T$. Such behaviour has been reported before~\cite{ding2009entanglement}.

\begin{figure}[H]
\begin{minipage}{15.5pc}
\includegraphics[width=18.1pc]{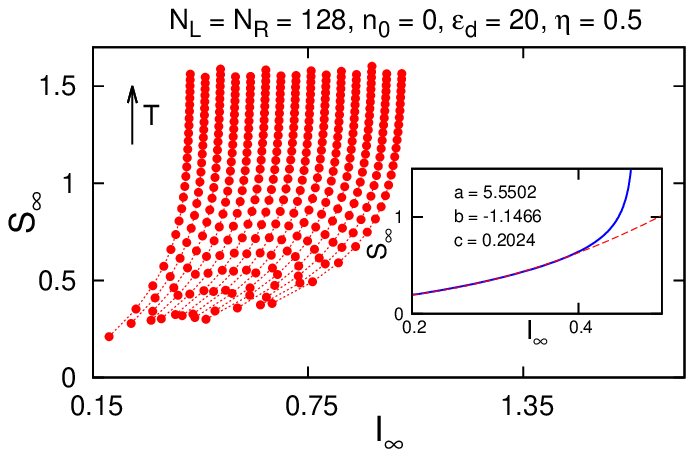}
\end{minipage}\hspace{1.7pc}
\begin{minipage}{15.5pc}
\includegraphics[width=18.1pc]{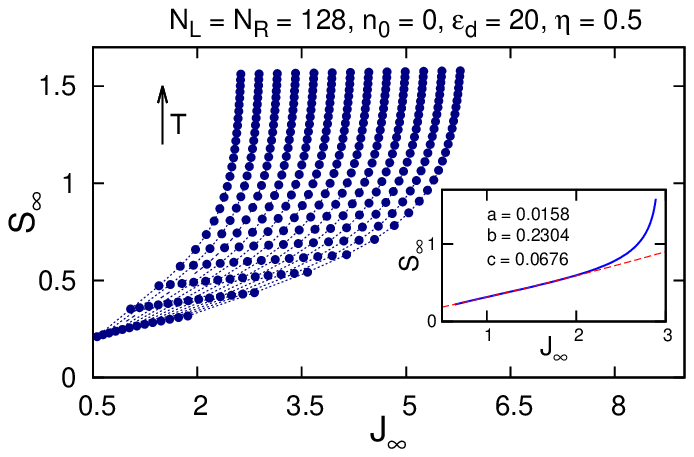}
\end{minipage}
\caption{\label{label7}The relationship between the steady state value of mutual information and current (particle $I_\infty$ and heat $J_\infty$). Inset shows the fitting of one curve to quadratic form $ax^2+bx+c$. Average temperature $T = \frac{T_{L}+T_{R}}{2}$ increases along the pointed direction.}
\end{figure}
We now proceed to study the behaviour of $S_{\infty}$ with respect to
$I_{\infty}$ and $J_{\infty}$. Fig.~\ref{label7} are obtained by
noting the steady state values of the current (particle and heat) and the mutual
information for different temperature differences. Each curve
corresponds to a particular $\Delta T$.  When the temperature is low
and the current depends explicitly on the temperature, we see by the
fit that the $S_{\infty}$ goes quadratically as a function of
$I_{\infty}/J_{\infty}$. In the limit of large temperatures, the current saturates to the
highest possible value but the mutual information keeps on increasing. Here we have shown data for just one representative sample of parameters $\epsilon_d=20,\eta = 0.5$, where a clean quadratic relationship between mutual information and current is obtained. 

\section{Summary and conclusions}
We studied transport through a bosonic dot as well as the dynamics of correlations between the dot and the baths by means of mutual information. Exploiting the symmetries present in the bath energies and the couplings, we were able to cast the expression for current in Landauer form. Analogously with the fermionic version~\cite{auditya2015landauer}, the similarities in the
current and the mutual information motivated us to further study the relationship between them. We found that with a suitable choice of the parameters, the steady state mutual information $S_{\infty}$ depends quadratically on the steady state
current $I_{\infty}$ and $J_{\infty}$, particularly when the initial
temperatures $T_{L}$ and $T_{R}$ are low. We also found that this
steady state value of mutual information varies logarithmically with
the total number of bosons present in the system. Our findings here are for a simple
noninteracting bosonic model, and it would be interesting to investigate
how general these relationships can be. We are currently in the process of
studying a more ``realistic" Hamiltonian of bosons, that has
position-position coupling giving rise to pairing terms. Also in progress is work studying other parameter regimes, where oscillatory steady states are found, and where the relationship between current and mutual information is less systematic.

\section*{Acknowledgement}
We thank Eran Rabani for helpful discussions. A.S is grateful to SERB for the startup grant (File Number: YSS/2015/001696).
\section*{References}
\bibliography{refs2}

\end{document}